\documentclass[twocolumn,showpacs,amsmath,aps,pra,amssymb,nofootinbib]{revtex4-1} 
\usepackage[T1]{fontenc}
\usepackage[latin9]{inputenc}
\usepackage{times}
\usepackage{color} 
\usepackage{xspace}
\usepackage{amssymb,amsmath}
\usepackage{amsbsy}
\usepackage{graphicx}
\usepackage{bm}
\usepackage{epstopdf}
\usepackage{float}
\usepackage{booktabs}

\usepackage[unicode,breaklinks]{hyperref}
\hypersetup{
    unicode=true,
    a4paper=true,
    plainpages=false, 
    colorlinks=true,
    linkcolor=blue,
    citecolor=blue,
    filecolor=black,
    urlcolor=blue
}
\newcommand{\bk}{\mathbf{k}}

\newcommand{\br}{\mathbf{r}}

\newcommand{\brho}{{\bm{\rho}}}

\newcommand{\UD}{U_{\mathrm{dd}}}

\begin{document}
 
\title{Thermally activated local collapse of a flattened dipolar condensate}  
 \author{E. B. Linscott}
 \author{P. B. Blakie}
\affiliation{Jack Dodd Centre for Quantum Technology, Department of Physics, University of Otago, Dunedin, New Zealand}
\begin{abstract} 
We consider the metastable dynamics of a flattened dipolar condensate. We develop an analytic model that quantifies the energy barrier to the system undergoing local collapse to form a density spike. We also develop a stochastic Gross-Pitaevskii equation (SGPE) theory for a flatted dipolar condensate, which we use to perform finite temperature simulations verifying the local collapse scenario. We predict that local collapses play a significant role in the regime where rotons are predicted to exist, and will be an important consideration for experiments looking to detect these excitations. 
\end{abstract}
 
 \maketitle 

\section{Introduction}
Tremendous recent progress with trapping and cooling highly magnetic atoms has enabled the production of dipolar Bose-Einstein condensates (BECs) \cite{Griesmaier2005a,Beaufils2008a,Mingwu2011a,Aikawa2012a}. In these condensates the atoms interact via an appreciable magnetic dipole-dipole interaction (DDI) that is both long-ranged and anisotropic, opening up a number of new many-body phenomena for exploration \cite{Baranov2008,Lahaye_RepProgPhys_2009}. 
 
A flattened dipolar condensate is produced by applying tight external confinement along one direction, and can be used to stabilize the system against the attractive component of the dipolar interaction \cite{Koch2008a,Muller2011a}. Novel predictions for dipolar condensates in this regime include density oscillating ground states \cite{Ronen2007a,Lu2010a,Asad-uz-Zaman2010a,Martin2012a},  roton-like excitations \cite{Santos2003a,Ronen2007a,Nath2010a,Hufnagl2011a,Blakie2012a,Corson2013a,JonaLasinio2013,Bisset2013a,Bisset2013b,Fedorov2014a}, modified collective and superfluid properties \cite{Wilson2010a,Ticknor2011a,Bismut2012a}, and stable 2D bright solitons  \cite{Pedri2005a}.  Many of these predictions require having a condensate in the dipole-dominated regime, i.e.~where the DDI is stronger than the short ranged contact interaction. Theoretical studies of this regime have mainly focussed on the elementary excitation spectrum, which can be calculated using Bogoliubov theory. However, density fluctuations in this regime can be large \cite{Blakie2013a,Bisset2013a,Baillie2014a} and recent work has shown that  Bogoliubov theory may be quite limited in applicability, particularly at finite temperature  \cite{Boudjemaa2013a}. 

To date, experiments in the flattened system have focused on quantifying the stability boundary \cite{Koch2008a,Muller2011a}, which can be explored by reducing the contact interaction (using Feshbach resonances) until the condensate becomes unstable. Theoretical work suggests that as the condensate crosses the stability boundary it undergoes a local collapse, in which it breaks up into a set of sharp density peaks \cite{Bohn2009a,Wilson2009a}  (also see \cite{Parker2009a}).
 
In this paper we show that a dipolar BEC is metastable against  local collapses even far from the stability boundary.  To do this we develop an analytic model in which we consider sharp density spikes (i.e.~a local collapse) forming on top of a condensate.  This enables us to quantify the energy barrier to  collapse. We then introduce a finite temperature dynamical model for the system by extending the SGPE formalism \cite{cfieldRev2008} to include DDIs. Our simulations with the SGPE demonstrate thermally activated local collapse events and support our density spike model.  
Our results indicate that metastability effects will be an important consideration for experiments aiming to verify the array of predictions that have been made for dipolar condensates in the flattened regime, such as the emergence of roton-like excitations.

\section{Model}\label{Sec:model}
\subsection{Uniform ground state}\label{S:BGGPE}
We consider a dipolar BEC that is harmonically confined along the $z$ direction and unconfined in the radial plane. 
The condensate wave function $\psi_0$ satisfies the non-local Gross-Pitaevskii equation (GPE)
\begin{equation}
\mu\psi_0(\br)=\left[h_{\mathrm{sp}}+\int d\br'U(\br-\br')|\psi_0(\br')|^2\right]\psi_0(\br),\label{e:fullGPE}
\end{equation}
where $\mu$ is the chemical potential and
\begin{equation}
 h_{\mathrm{sp}}=-\frac{\hbar^2\nabla^2}{2m}+\frac{m\omega_z^2z^2}{2},
\end{equation}
is the single particle Hamiltonian, with $\omega_z$ being the axial trap frequency and $m$ the atomic mass.

The atoms we consider are taken to have an appreciable magnetic dipole momentum $\mu_m$ polarized along the $z$-axis by an external magnetic field. 
In this case the associated interaction potential is $
\UD(\br)=\frac{3g_d}{4\pi}{[1-3(\hat{\mathbf{z}}\cdot\hat{\mathbf{r}})^2]}/{r^3},$
where $g_d=\mu_0\mu_m^2/3$ is the DDI coupling constant and $\hat{\mathbf{r}}=\mathbf{r}/|\mathbf{r}|$. The particles can also interact by a short ranged contact interaction with coupling constant $g_s=4\pi a_s\hbar^2/m$, where $a_s$ is the scattering length, so that the full interaction is $U(\br)=g_s\delta(\br)+U_{\mathrm{dd}}(\br)$ (e.g.~see \cite{Yi2000a,Yi2001a,Lahaye_RepProgPhys_2009}).

The condensate solution to Eq.~(\ref{e:fullGPE}) takes the form $\psi_0(\br)=\sqrt{n_0}\chi_\sigma(z)$, where $n_0$ is the areal density, and $\chi_\sigma$ is a normalized axial mode. Here we approximate $\chi_\sigma$ as a Gaussian of the form
  \begin{equation}
  \chi_\sigma(z)=\frac{1}{\pi^{1/4}\sqrt{\sigma l_z}}e^{-z^2/2\sigma^2 l_z^2},
  \end{equation}
  with length scale $l_z=\sqrt{\hbar/m\omega_z}$. We treat $\sigma$ as a variational parameter to be determined by minimizing the energy functional
  \begin{align}
      E[\psi]&=\int\!d\br\,\psi^*(\br)\left[ h_{\mathrm{sp}}+\frac{1}{2}\int d\br'U(\br-\br')|\psi(\br')|^2\right]\psi(\br),\label{e:Efun}
\end{align}
which, upon substituting the Gaussian ansatz, gives
\begin{equation}
E_\sigma=n_0A\hbar\omega_z\left[\frac{1}{4\sigma^2}+\frac{\sigma^2}{4}+\frac{\nu_s+2\nu_d}{2\sqrt{2\pi}\sigma}\right].\label{e:Esigma}
\end{equation}
Here $A$ is the area of the system and we have introduced $\nu_s=n_0g_s/\hbar\omega_zl_z$ and  $\nu_d=n_0g_d/\hbar\omega_zl_z$ as the dimensionless contact and DDI parameters, respectively.
For $|\nu_s+2\nu_d|\ll1$ the minimum value of $\sigma$ approaches  $1$, i.e.~the quasi-2D regime \cite{Fischer2006a}. In general the variational Gaussian approach we use here has been shown to provide an accurate description even for  large interaction parameter values \cite{Baillie2014b}. Using the value of $\sigma$ that minimizes Eq.~(\ref{e:Esigma}), the condensate chemical potential  [c.f.~Eq.~(\ref{e:fullGPE})] is given by
\begin{equation}
\mu_\sigma =  \hbar\omega_z\left[\frac{1}{4\sigma^2}+\frac{\sigma^2}{4}+\frac{\nu_s+2\nu_d}{\sqrt{2\pi}\sigma}\right].\label{e:Musigma}
\end{equation}

\subsection{Density spike model}\label{S:peakmodel}
\begin{figure}[ht!] 
   \centering
   \includegraphics[width=3.30in]{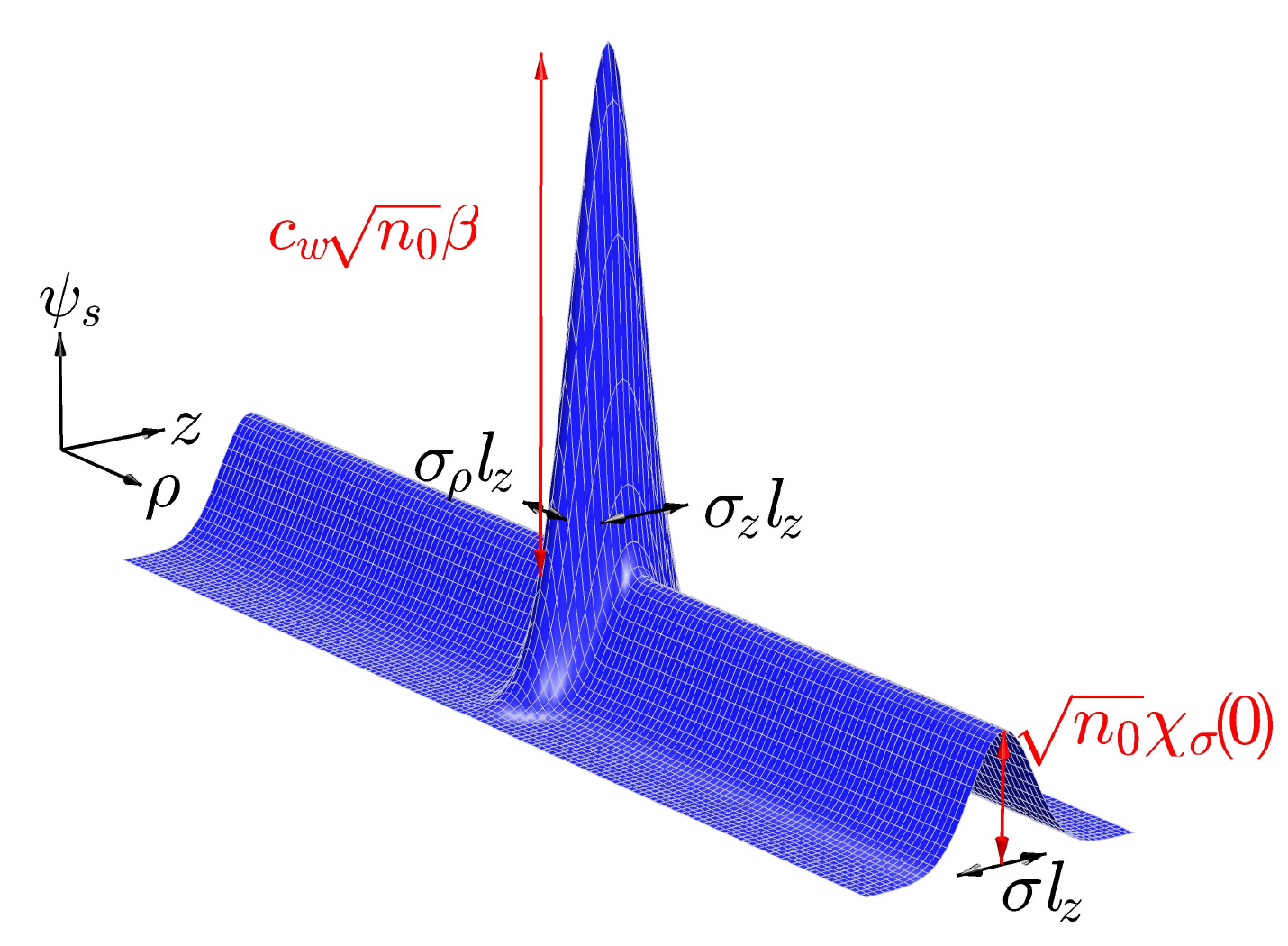}  
   \caption{(colour online) Visualisation of the density spike ansatz [see Eq.~(\ref{e:model})]  illustrating the  parameters used, with $c_w=(\pi^{3/4}\sigma_\rho\sqrt{\sigma_zl_z})^{-1}$.  }
   \label{fig:pimple}
\end{figure}

We want to consider the energetics of the system forming density spikes on top of the flat condensate ground state. To do this we propose a variational ansatz for a condensate with a Gaussian density spike of the form 
\begin{equation}
\psi_s(\br)=\sqrt{n_0}\chi_\sigma(z)+\sqrt{n_0}\beta\frac{\exp\left[-\frac{1}{2}\left(\frac{z^2}{\sigma_z^2l_z^2}+\frac{\rho^2}{\sigma_\rho^2l_z^2}\right)\right]}{\pi^{3/4}\sigma_\rho \sqrt{\sigma_z l_z}},\label{e:model}
\end{equation}
where $\bm{\rho}=(x,y)$ is the in-plane coordinate and the last term describes the spike in terms of dimensionless height $\beta$ and width parameters  $\{\sigma_\rho,\sigma_z\}$ (see Fig.~\ref{fig:pimple}). 

We consider a large system, so that a single spike has negligible effect on the condensate itself. Consequently, we take the condensate variational parameter $\sigma$ to be determined by minimizing Eq.~(\ref{e:Esigma}) irrespective of the peak (and hence $\sigma$ is a function of $\nu_s+2\nu_d$ only).

\begin{figure}[ht!] 
   \centering
   \includegraphics[width=3.2in]{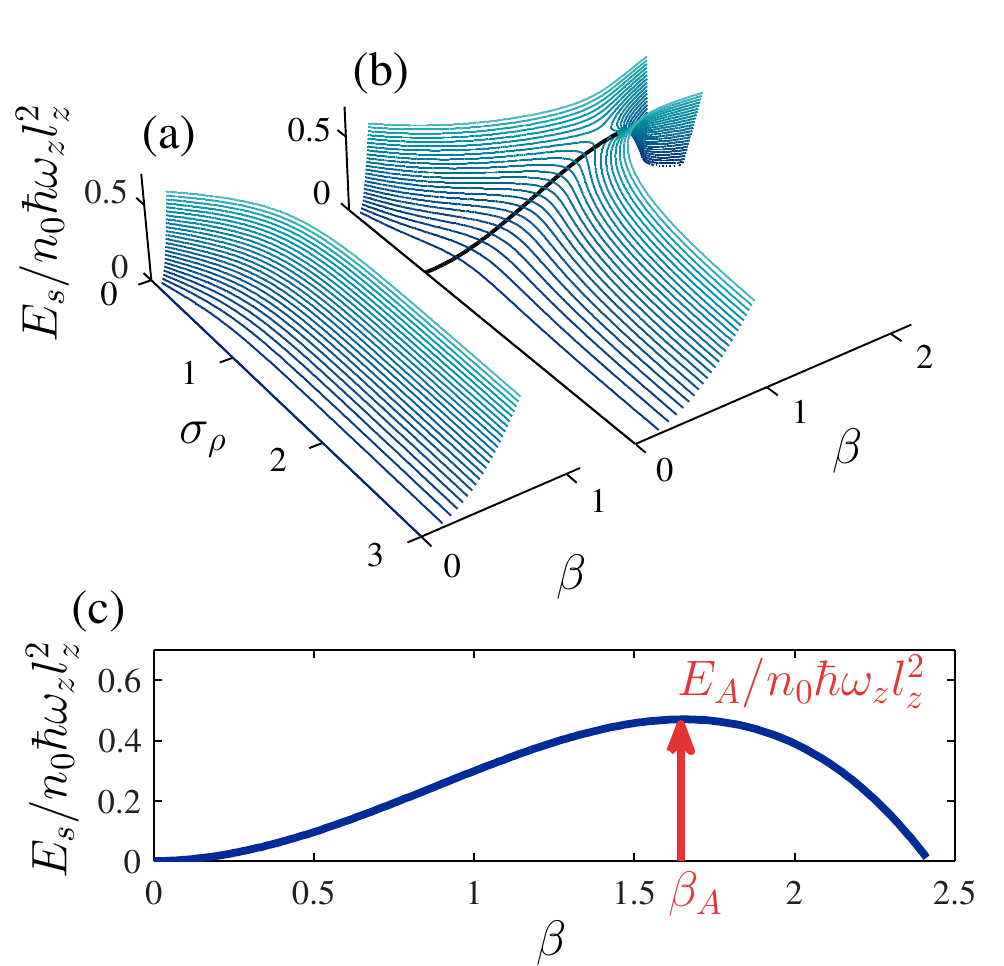}  
   \caption{(colour online) Spike formation energy surface $E_s(\beta,\sigma_\rho,\sigma_z)$. Results shown as a function of $\{\sigma_\rho,\beta\}$ for (a) stable regime $\nu_d<\nu_s$ , with $\nu_d = 0.75$, $\nu_s = 1$ and (b) metastable regime $\nu_d>\nu_s$, with  $\nu_d = 1.4$, $\nu_s = -0.3$. In (a) we set $\sigma_z = \sigma = 1.22$ for simplicity. In (b), we choose $\sigma_z = 1.35$, which minimizes the activation energy $E_A$. (c) Spike energy crossing the saddle of the energy surface  along path shown in (b). Activation energy $E_A$ and the value of $\beta$ at the activation point ($\beta_A$) are indicated.}
   \label{fig:esurf}
\end{figure}

The energy associated with forming a peak on top of a condensate background is then evaluated by substituting (\ref{e:model}) in Eq.~(\ref{e:Efun}), which yields
\begin{widetext}
\begin{align}
E_s\equiv&  E[\psi_s]-E[\psi_0]-\mu_\sigma N_s,\nonumber \\
 =&n_0 l_z^2 \hbar \omega_z \left\{2\sqrt{2 \pi}\beta \sigma_\rho \left(\frac{\sigma \sigma_z}{\sigma^2+\sigma_z^2}\right)^{3/2}\left(\sigma \sigma_z + \frac{1}{\sigma \sigma_z} \right)+\frac{\beta^2}{2}\left(\frac{\sigma_z^2}{2} + \frac{1}{2\sigma_z^2} + \frac{1}{\sigma_\rho^2}\right)\right. -\mu_\sigma\left(4\sqrt{2\pi}\beta\sigma_\rho\sqrt{\frac{\sigma\sigma_z}{\sigma^2+\sigma_z^2}}+\beta^2\right)\nonumber \\ 
&\hspace{1.5cm}+\frac{4\beta \sigma_\rho (\nu_s+2\nu_d)}{\sqrt{\frac{3}{2}\sigma\sigma_z+\frac{1}{2}\sigma^3/\sigma_z}}
+\frac{3 \beta^2}{\sqrt{\pi(\sigma^2+\sigma_z^2)}}\left(\nu_s+  \frac{2}{3}\nu_d\left[1+f\left(\frac{\sqrt{\sigma^2+\sigma_z^2}}{\sigma}\frac{\sigma_\rho}{\sigma_z}\right)\right]\right)\nonumber \\ 
&\left.\hspace{2cm} +\frac{4\beta^3}{3\pi\sigma_\rho\sqrt{\frac{3}{2}\sigma \sigma_z+\frac{1}{2}\sigma_z^3/\sigma}}\left[\nu_s+ \nu_df\left(\sqrt{\frac{\sigma^2+\sigma_z^2}{\sigma^2+\frac{1}{3}\sigma_z^2}}\frac{\sigma_\rho}{\sigma_z}\right)\right]
+\frac{\beta^4}{2(2\pi)^{3/2}\sigma_z \sigma_\rho^2}\left(\nu_s +\nu_d f(\sigma_\rho/\sigma_z)\right)\right\}
\end{align}
\end{widetext}
where 
\begin{align}
f(\kappa) &\equiv   \frac{2\kappa^2 + 1}{\kappa^2 - 1}-\frac{3\kappa^2\arctan\left(\sqrt{\kappa^2 - 1}\right)}{\left(\kappa^2 - 1\right)^{3/2}}
\end{align}
is a monotonically increasing function of $\kappa$ with $f(0)=-1$ and $f(\infty)=2$ \cite{Giovanazzi2003a}.
The term $\mu_\sigma N_s$ accounts for the energy liberated by removing atoms from the condensate to form the spike, where the number of atoms in the spike is  
\begin{align}
N_s&\equiv\int d\br (|\psi_s|^2-|\psi_0|^2)\nonumber \\
&= n_0 l_z^2\beta\!\left(4\sqrt{2\pi} \sqrt{\frac{\sigma \sigma_z \sigma_\rho^2}{\sigma^2+\sigma_z^2}}+\beta\right)\!.
\end{align} 
Some examples of the spike energy  $E_s(\beta,\sigma_\rho,\sigma_z)$ are presented in Fig.~\ref{fig:esurf}. For $\nu_s>\nu_d$  [Fig.~\ref{fig:esurf}(a)] the dipolar condensate is stable, in that the energy cost of forming a density spike is positive and increases with increasing $\beta$. In contrast for the dipole dominant regime $\nu_d>\nu_s$ [Fig.~\ref{fig:esurf}(b)] the condensate is metastable: the energy can be lowered by the formation of a dense narrow spike. However, spikes of intermediate densities still cost energy, presenting a barrier to the formation of a high density spike. We note that our formalism will be invalid for an extremely dense spike, but is adequate for quantifying the properties of the energy barrier and the system's passage over it.

In Fig.~\ref{fig:esurf}(b) we indicate a path along which a high density peak might form. This path crosses the energy barrier at its lowest point, with the value of the energy along this path shown in Fig.~\ref{fig:esurf}(c). We define the minimum height of the energy barrier [at the saddle point of the function $E_s(\beta,\sigma_\rho,\sigma_z)$] as the \textit{activation energy} $E_A$, and label the associated value of $\beta$ at this point as $\beta_A$, corresponding to a peak areal density of 
\begin{equation}
n_A = n_0\left(1+\frac{2\beta_A}{\pi^{1/2}\sigma_\rho}\sqrt{\frac{2 \sigma \sigma_z}{\sigma^2+\sigma_z^2}}+\frac{\beta_A^2}{\pi \sigma_\rho^2}\right).
\end{equation}
\begin{figure}[hb!] 
   \centering
   \includegraphics[width=3.40in]{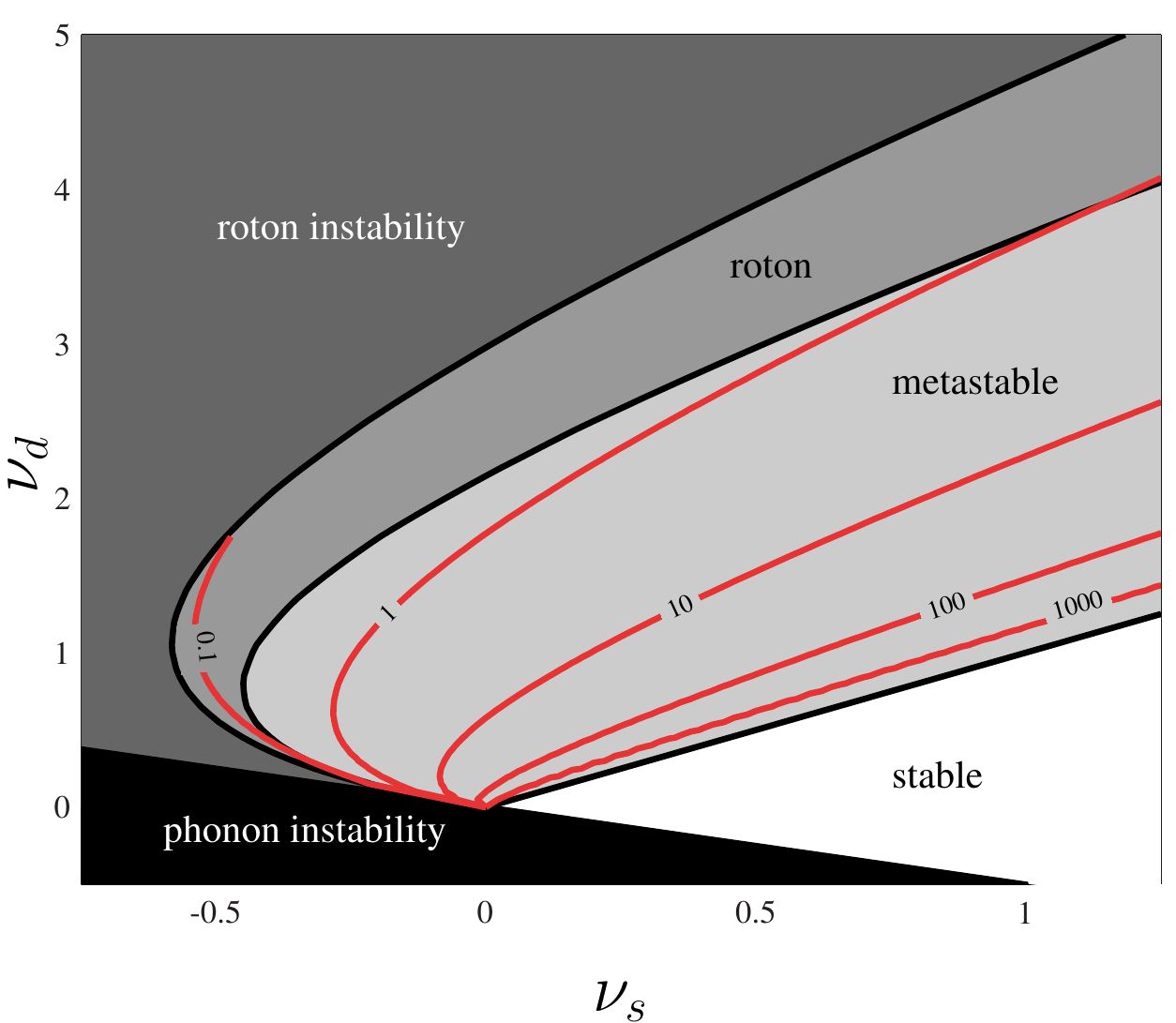}  
   \caption{(colour online)  Phase diagram and metastable energy barrier.  The stable, metastable regimes (which includes the roton regime), and regions of instability are indicated.  
   Contours indicate values of the energy barrier ${E}_A$ in units of $n_0l_z^2\hbar\omega_z$.   }
   \label{fig:EA}
\end{figure}
The activation energy varies as a function of the dimensionless interaction parameters $\nu_s$ and $\nu_d$, and contours of this are shown in Fig.~\ref{fig:EA}. For reference we have placed these contours on top of a stability diagram for the system, obtained by examining the behaviour of the condensate quasiparticles as a function of their in-plane wave vector $k_\rho$ (see \cite{Santos2003a,Blakie2012a,Baillie2014b} for additional discussion of these regimes).
Notably a number of stable and unstable regions can be identified by the quasiparticle spectrum: In the  \textit{phonon instability} region a long wavelength  ($k_{\rho}\to0$) quasiparticle becomes dynamically unstable (i.e.~its energy becomes imaginary). In the \textit{roton instability} region a short wavelength quasiparticle (i.e.~$k_{\rho}\sim1/l_z$)  is dynamically unstable. The \textit{metastable region}  occurs when interactions are dipole-dominated $\nu_d>\nu_s$ and all the quasiparticles have real positive energies. It is denoted as metastable because, as quantified by our model,  the condensate is nevertheless able to lower its energy by forming density spikes, even though this is not revealed in the quasiparticle spectrum. The \textit{roton} region is part of the metastable region, and occurs when the dispersion relation has a  roton-like feature i.e.~a local minimum at non-zero $k_{\rho}$.

The  results of Fig.~\ref{fig:EA} indicate that in the regime where rotons occur the activation energy $E_A$ is typically quite low, so that we would expect  density spikes to form via thermal activation or tunneling.  The results also show that in the roton regime and for larger values of $\nu_s$,  the activation energy increases.

We note that for $\nu_d=-\frac{1}{2}\nu_s$ (i.e.~the upper boundary of the phonon instability region) the effective long wavelength interaction [c.f~last term in Eq.~(\ref{e:Esigma})] is zero, and $E_A$ approaches 0.
For the case $\nu_d<-\frac{1}{2}\nu_s$  the  effective long wavelength interaction is attractive and the condensate unstable to a long-wavelength phonon collapse. It is worth noting that within this regime it has been predicted that stable bright solitons should exist (e.g.~see \cite{Pedri2005a}).

\section{SGPE simulations}
To verify and explore the local instability predicted by our Gaussian ansatz, we now proceed to consider a finite temperature dynamical description of a planar dipolar condensate, based on the  SGPE formalism.

\subsection{SGPE theory for planar dipolar BEC}
The SGPE formalism treats the thermal dynamics of the low energy modes of a partially condensed Bose field. Essentially the formalism provides a classical field (i.e.~Gross-Pitaevskii-like evolution) for the low energy modes, with additional damping and noise terms to describe the coupling to high energy (non-classical) modes of the system (e.g.~see \cite{Stoof2001a,Gardiner2003a,Bradley2008a,cfieldRev2008,Proukakis2008a}).

The SGPE evolution of this system is given by
\begin{align}
 d\Psi = \mathcal{P}\!\left\{-\frac{(i+\gamma)}{\hbar}(\mathcal{L}-\mu)\Psi\,dt+\!\sqrt{2\gamma k_BT/\hbar}\,dW(\brho)\!\right\}\!,\label{e:SGPE}
\end{align}
where $\Psi=\Psi(\brho)$ is the quasi-2D classical field for the system, with $\brho=(x,y)$,
\begin{align}
\mathcal{L}\Psi &= -\frac{\hbar^2\nabla_{\brho}^2}{2m}\Psi+\mathcal{F}^{-1}_{\brho}\left\{\tilde{U}_{\mathrm{2D}}(\bk_{\rho})\mathcal{F}_{\brho}\{|\Psi(\brho)|^2\}\right\}\Psi,\label{e:LGPE}
\end{align}
is the effective 2D Gross-Pitaevskii operator and $\mathcal{F}_{\brho}$ is the in-plane Fourier transform. To obtain this form we have integrated out the $z$-dimension, resulting in the effective 2D interaction potential in $k_\rho$-space
\begin{align}
\tilde{U}_{\mathrm{2D}}(\bk_{\rho}) &\equiv \int dk_z \tilde{U}(\bk) \mathcal{F}_z\left\{|\chi_\sigma(z)|^2\right\}, \\
&=\frac{1}{\sqrt{2\pi} l_z}\left[g_s+g_d(2-3\sqrt{\pi}Qe^{Q^2}\mathrm{erfc}\,Q)\right]
\end{align}
where $Q=k_\rho l_z/\sqrt{2}$. The stochastic term $dW$ is a complex Gaussian noise satisfying $\langle dW\rangle=\langle dW^2\rangle=0$, $\langle dW(\brho)dW^*(\brho')\rangle=\delta(\brho-\brho')dt$. In Eq.~(\ref{e:SGPE}) a projector $\mathcal{P}$ appears which is used to restrict the evolution to the low energy appreciably occupied modes of the field. Because we consider a uniform planar system this is implemented as a radially symmetric cutoff $k_{\mathrm{cut}}$ in wave-vector space, i.e.~the low energy region evolved is restricted to parts of $\Psi$ with $|\bk_\rho|<k_{\mathrm{cut}}$. 

The parameter $\gamma$ describes the coupling to high energy modes (treated as a reservoir at temperature $T$ and chemical potential $\mu$) that have been eliminated from  $\Psi$ by the projector. For the case of contact interactions $\gamma\sim (a_s/\lambda_{\mathrm{dB}})^2$, where $\lambda_{\mathrm{dB}}=h/\sqrt{2\pi mk_BT}$ \cite{Rooney2012a}.  A detailed microscopic derivation of the SGPE theory along the lines of \cite{Gardiner2003a} has not been performed for the case of a planar dipolar gas, however the theory is phenomenologically justified for our purposes of studying dynamics near equilibrium: the SGPE theory is a Langevin equation that provides a grand-canonical classical field description of the low energy modes of the field, with the damping (being the term in (\ref{e:SGPE}) proportional to $\gamma$) and noise (the term proportional to $\sqrt{\gamma}$) being related through the fluctuation dissipation theorem\footnote{It is worth noting that equilibrium properties are independent of $\gamma$.}.

In formulating the SGPE theory for the planar system we have made the quasi-2D approximation, so that all motion in the $z$-direction is frozen in the harmonic oscillator ground state.

\subsection{Simulations}
\subsubsection{Uniform simulation scheme}
We perform our simulations of Eq.~(\ref{e:SGPE}) on a square domain of  area $A=L\times L$, where $L$ is the side length,  and subject to periodic boundary conditions. 
The classical field can therefore be represented effectively in a plane wave basis,
\begin{equation}
\Psi(\brho,t)=\sum_{\bk_\rho} c_{\bk_\rho}(t)\frac{e^{i {\bk_\rho} \cdot \brho}}{\sqrt{A}},
\end{equation}
where the in-plane wave vectors are ${\bk_\rho}=2\pi(n_x,n_y)/L$, $n_x,n_y\in\mathbb{Z}$, and the $c_{\bk_\rho}$ are complex time-dependent amplitudes.  The numerical scheme used to simulate the SGPE is the 2D version of the fast Fourier transform-based algorithm discussed in Sec.~III of Ref.~\cite{Blakie2008a}, with an additional step introduced to evaluate the convolution involving the $\bk$-dependent interaction [see Eq.~(\ref{e:LGPE})].

\subsubsection{Initial condition}
For our initial condition we sample a randomized state constructed from a  condensate and Bogoliubov quasiparticles according to
\begin{align}
\Psi(\brho,0)=\sqrt{n_0}\!+\!\sum_{\bk_\rho} \left(u_{\bk_\rho} \alpha_{\bk_\rho} - v_{-\bk_\rho} \alpha_{-\bk_\rho}^*\right)\frac{e^{i {\bk_\rho} \cdot \brho}}{\sqrt{A}} ,
\end{align}
where   $\alpha_{\bk_\rho}=
\sqrt{\frac{k_B T}{2 \epsilon_{\bk_\rho}}}(u_r+iu_i)$, with 
 $u_r$ and $u_i$ being normally distributed random numbers generated for every ${\bk_\rho}$. 
 In the above expression we have introduced the Bogoliubov quasiparticle  energy $\epsilon_{\bk_\rho}$ and amplitudes $\{u_{\bk_\rho},v_{\bk_\rho}\}$, which are
 \begin{align}
 \epsilon_\bk&=\sqrt{\frac{\hbar^2k_\rho^2}{2m}\left[\frac{\hbar^2k_\rho^2}{2m}+2n_0\tilde{U}_{\mathrm{2D}}({\bk_\rho})\right]},\\
u_{\bk_\rho}&=\sqrt{\frac{1}{2}\left(\frac{\frac{\hbar^2{k_\rho}^2}{2m}+ n_0\tilde{U}_{\mathrm{2D}}({\bk_\rho})}{\epsilon_{\bk_\rho}}+1\right)},\\
v_{\bk_\rho}&=\sqrt{\frac{1}{2}\left(\frac{\frac{\hbar^2{k_\rho}^2}{2m}+ n_0\tilde{U}_{\mathrm{2D}}({\bk_\rho})}{\epsilon_{\bk_\rho}}-1\right)}\mathrm{sign}\left[\tilde{U}_{\mathrm{2D}}({\bk_\rho})\right].
 \end{align}
This choice of initial state ensures that every quasiparticle mode is occupied according to the classical limit of the Bose-Einstein distribution, and we find that it changes little when allowed to equilibrate via the SGPE.
 
 \subsubsection{Simulation parameters}
For the simulations we present we take $L=80\,l_z$  and  use a cutoff momentum of $k_{\mathrm{cut}}=\sqrt{10}/l_z$. For this choice $ 5097$ plane wave modes are retained in classical region for which the dynamics are simulated. We focus on the case of a condensate of density  $n_0=4/l_z^2$, with interaction parameters $\nu_s=-0.301$, $\nu_d=1.404$, which is in the metastable regime, with $E_A=3.28\,\hbar\omega_z$, $\beta_A = 1.54$.
The SGPE simulations are performed using reservoir parameters $\mu=\hbar \omega_z$ and  temperatures  in the range 0.2 to 0.45 $\hbar \omega_z/k_B$. We find that the condensate fraction of the field $\Psi$ varies from about $0.95$ at $T=0.2\hbar\omega_z/k_B$ to $0.88$ at $T=0.45\hbar\omega_z/k_B$.  The results we present are for the case of  $\gamma=0.1$.

\subsection{SGPE results}
\subsubsection{Observed dynamics}
\begin{figure}[ht!] 
   \centering
   \includegraphics[width=3.40in]{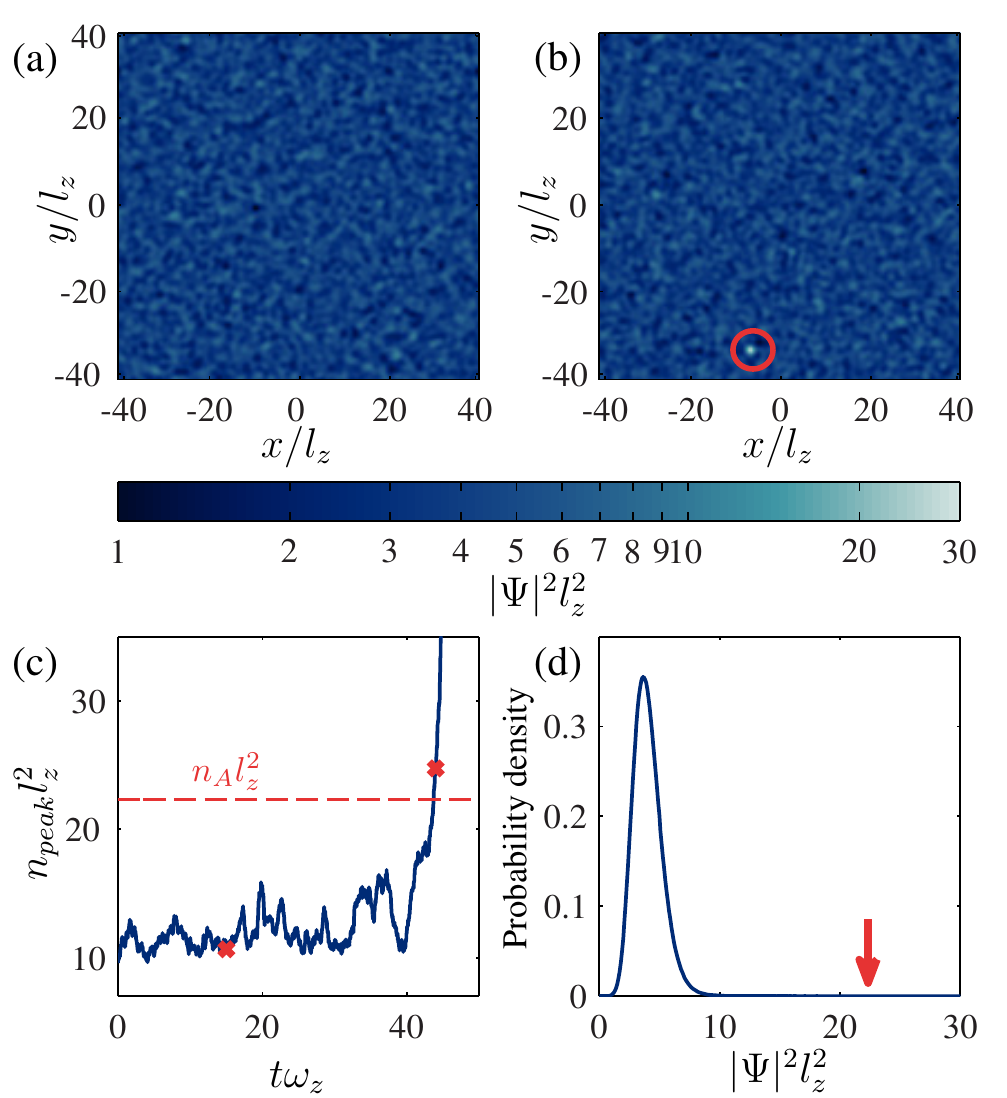}  
   \caption{ (color online) Field density and a typical spike formation event. The field density $|\Psi|^2$ is shown (a) at $t=15/\omega_z$ (prior to spike formation) and (b) at $t=44/\omega_z$ (during spike formation). The red circle indicates the spike location. (c) The peak density in the system during the simulations, revealing the sudden formation of a spike at $t\approx45/\omega_z$. The red crosses indicate the two times corresponding to the fields plotted in (a) and (b).   (d) The distribution of densities across the simulation cell prior to collapse. 
   The red arrow indicates $n_A = 22.3/l_z^2$. The simulation parameters were $T = 0.2\,\hbar\omega_z/k_B$,  $\nu_s=-0.301$, and $\nu_d=1.404$.
   }
   \label{fig:sim_data}
\end{figure}

An example of the density profile during a typical SGPE evolution is shown in Fig.~\ref{fig:sim_data}(a). The noisy density pattern reveals the fluctuating thermal modes in the low energy region, and is similar to the typical results of SGPE evolution in the case of contact interactions (e.g.~see Fig.~2 of \cite{Davis2002a}). However, for this dipolar simulation in the metastable regime, we eventually find that a density spike emerges [see Fig.~\ref{fig:sim_data}(b)], which persists in the field. It is useful to define the instantaneous peak density of the field
\begin{equation}
n_{\mathrm{peak}}(t)=\max_{\boldsymbol{\rho}}\left\{|\Psi(\boldsymbol{\rho},t)|^2\right\},
\end{equation}
i.e.~as the maximum density occurring  at any grid point. In Fig.~\ref{fig:sim_data}(c) we quantify the behaviour of $n_{\mathrm{peak}}$ in the evolution leading up to the density spike forming: this formation is clearly revealed by the sudden onset of rapid growth of $n_{\mathrm{peak}}$ at  $t\approx45/\omega_z$. To put these values of peak density into context, in Fig.~\ref{fig:sim_data}(d) we show the probability density function for values of density occurring in the field. This is obtained by making a histogram of the density values occurring at every grid point using the field sampled at a discrete set of times prior to the collapse. This density distribution revels that the most likely density is $\sim4/l_z^2 = n_0$. The thermal fluctuations in the field give rise to the spread in the distribution function around the most likely value, and we emphasize that the spike formation proceeds through values that are out in the tails of this distribution [as indicated in Fig.~\ref{fig:sim_data}(d)].

 The time it takes for a spike to form is stochastic and can vary significantly between different SGPE simulations for identical parameters. Spike formation times tend to get shorter the closer the system is to the roton instability boundary and as the temperature increases. Once formed, the spikes grow rapidly as shown in Fig.~\ref{fig:sim_data}(c). Overall these qualitative observations are consistent with the spikes occurring as a thermally activated crossing of the energy barrier consistent with our simple model of Sec.~\ref{Sec:model}.

\subsubsection{Characterizing spike formation}
\begin{figure}[ht!] 
   \centering
   \includegraphics[width=3.20in]{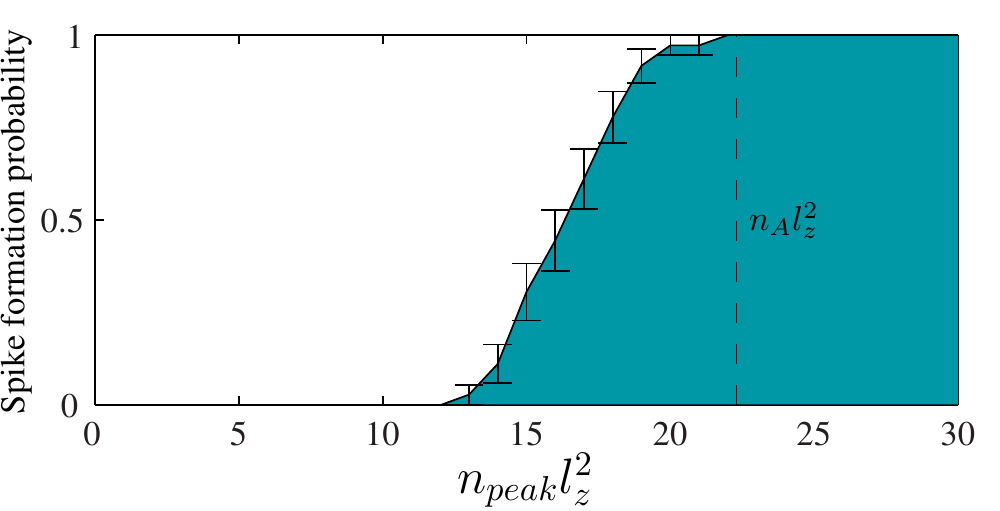}  
   \caption{(color online) The probability that a density spike forms within a time interval  of $\delta t=5/\omega_z$ after a particular peak density $n_{\mathrm{peak}}$ occurs in the simulation. Calculations for $T=0.2\,\hbar\omega_z/k_B$.}
   \label{fig:np}
\end{figure}

It is evident, particularly from Fig.~\ref{fig:sim_data}(c) and (d),  that spike formation is due to fluctuations in density to large values. We aim to measure the correlations between a peak density of some value occurring in the field and a spike forming. 
To do this we calculate the probability that a spike forms within a time interval of $\delta t=5/\omega_z$ after a value of $n_{\mathrm{peak}}$ occurs in in the field.  
We take $|\Psi|^2 > 30/l_z^2$ as an unambiguous measure of a spike having formed in the system, as this density was only ever observed to occur once a spike had formed and was growing rapidly.
The probability that a spike forms was then calculated using 36 trajectories of the SGPE for the parameters of  Fig.~\ref{fig:sim_data}  with the results  shown in Fig.~\ref{fig:np}. These indicate that if a density fluctuates to a value exceeding $\sim 16$  then a spike is   likely  to form. 
This is a lower, but comparable, value to the density at the activation point ($n_A = 22.3/l_z^2$) as predicted by our Gaussian model\footnote{This is the model discussed in Sec.~\ref{Sec:model}, but with  $\sigma =\sigma_z=1$, consistent with the quasi-2D restriction of the SGPE model.}. We also note that the typical widths of the observed spikes in the SGPE simulations are in quantitative agreement with the value of $\sigma_\rho$ predicted by the model at the activation point.

\begin{figure}[ht!]  
   \centering
   \includegraphics[width=3.2in]{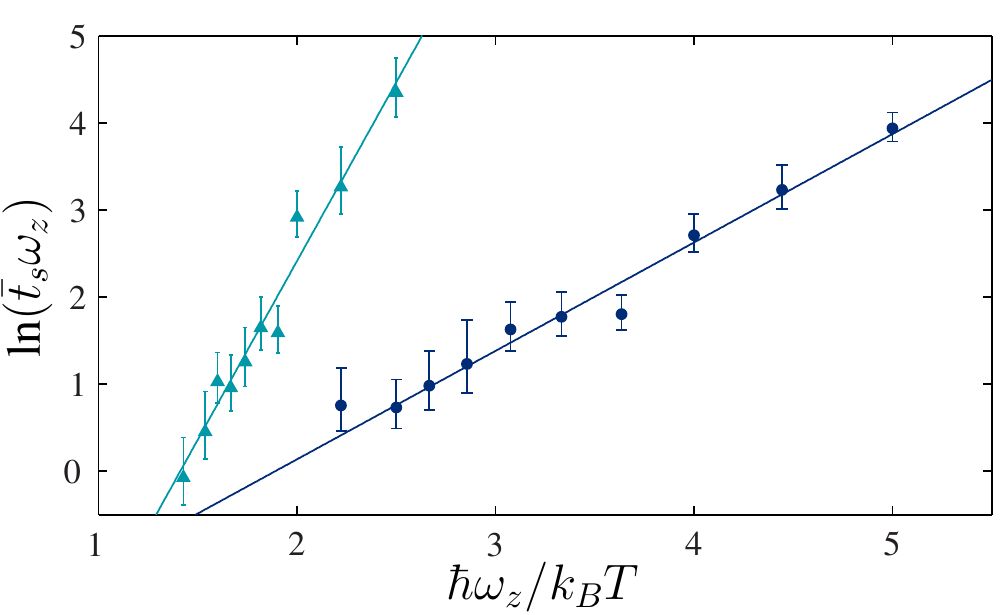}  
   \caption{ (color online) Temperature-dependence of the mean peak formation time $\bar t_s$, plotted here for two different sets of interaction parameters:  (circles) $\nu_s=-0.301$, $\nu_d=1.404$ (as in earlier results), and (triangles) $\nu_s= -0.201$,  $\nu_d= 1.354$. The linear fits have slopes of $1.25 \pm 0.09$ and $4.1 \pm 0.4$. }
   \label{fig:tvT}
\end{figure}

Finally, we consider the influence of temperature on the rate at which spikes form. We define the mean spike formation time $\bar{t}_s$ to be the average evolution time until a spike forms, and calculate it by averaging the individual times spike formation times obtained from 10 -- 20  SGPE simulations for each parameter set. We present results for the dependence of $\bar{t}_s$ in Fig.~\ref{fig:tvT} for two sets of interaction parameters, and for a range of temperatures. These results demonstrate that the mean spike formation time scales as $\bar{t}_s\sim\exp(c\hbar \omega_z/k_BT)$, which corresponds to  Arrhenius' scaling with temperature (e.g.~see \cite{GardinerStochMethods}), where we take $c$ to be a fit parameter. The fits to the SGPE results give $c = 1.25 \pm 0.09$ and  $4.1 \pm 0.4$. For comparison, the Gaussian model predicts activation energies of $E_A=3.28\hbar\omega_z$ and $E_A=5.51\hbar\omega_z$ respectively. Thus we see that as the metastable energy barrier increases, the rate of spike formation decreases.

We have not systematically studied the effect of changing $\gamma$, but in simulations where we reduced $\gamma$ by two orders of magnitude\footnote{In this small $\gamma$ limit the theory reduces to the so called projected-GPE theory or classical field method (see \cite{Pawowsk2013a}), providing a micro-canonical  description of the low energy system modes.} we found that the mean peak formation time was changed by about a factor of 2.

\section{Conclusion and Outlook}
In this paper we have considered the energetics and finite temperature dynamics of a flattened dipolar condensate. By developing an analytic model we show that it is energetically favorable for density spikes to form in this system in the metastable dipole-dominated regime, and we have characterized the energy barrier to formation as a function of the interaction parameters. Notably, our results predict that the role of local density spikes will be important in the regime where rotons are predicted to exist in the elementary excitation spectrum. Developing the SGPE theory for this system,  we have shown that thermal fluctuations can nucleate density spikes, and that their properties are consistent with our analytic model. The density spikes we discuss here realize a local collapse scenario   \cite{Bohn2009a}, whereby atoms far away from the spike remain unaffected (c.f.~global collapse for  condensates with attractive contact interactions \cite{Donley2001a}). Our theory here has only considered the formation dynamics of the spike, and does not provide a consistent model of the spike after it forms (and having passed beyond the energy barrier). It is likely that the atoms within the spike will be lost by three-body recombination (increased significantly due to the high density in the spike), and will lead to heating in the system. Because the number of atoms in a given spike is a small fraction of the system, the development of a single spike will not necessarily be detrimental to the condensate, and many such local collapses may be required to heat the condensate. Qualitatively, such a scenario seems consistent with the experiments of Koch \textit{et al.}~\cite{Koch2008a}. For example, in Fig.~2 of \cite{Koch2008a} a continuous decrease in the condensate number was observed as the stability boundary was approached. Indeed, this suggests that condensate lifetime measurements would be a possible avenue for experiments to investigate the energy barrier to local collapse in the dipole-dominated regime.

It is useful to put the parameters of our calculations into context of current experiments. The case considered in Fig.~\ref{fig:sim_data} corresponds to the central region of a  $55\times10^3$ atom $^{164}$Dy  condensate in a 3D harmonic trap with frequencies of $(f_\rho,f_z)=(15,10^3)$ Hz, and scattering length $a_s=-28\,a_0$, where $a_0$ is the Bohr radius. Translating the results of Fig.~\ref{fig:tvT} for this case (i.e.~the filled circle results) give that at temperatures of $10\,$nK the mean spike formation times $\bar t_s$ will be $\sim 6\,$ms, decreasing to $0.2\,$ms at $25\,$nK. That said, we emphasize that a precise model of the experimental regime will require accounting for the effects of radial trapping.

An important extension of the work in this paper will be to develop a more detailed analytic theory of the collapse dynamics. For example, the stochastic Lagrangian approach used in Ref.~\cite{Duine2001a} could be extended to the dipolar case.

\section*{Acknowledgments:}
We thank D.~Baillie for his assistance, and A.~S.~Bradley for useful discussions.
Support by the Marsden Fund of New Zealand (contract number UOO1220) is gratefully acknowledged.

 
%

\end{document}